\documentclass[11pt,twoside]{article}
\usepackage{asp2004}
\usepackage{psfig}
\usepackage{epsf}
\usepackage{graphics}
\usepackage{lscape}
\def\edcomment#1{\iffalse\marginpar{\raggedright\sl#1\/}\else\relax\fi}
\reversemarginpar

\begin{document}
\title{Be/X-ray binaries: An observational approach}
\author{I. Negueruela}
\affil{Dpto. de F\'{\i}sica, Ingenier\'{\i}a de Sistemas y Teor\'{\i}a de 
la Se\~{n}al, Universidad de Alicante, Apdo. 99, E03080 Alicante, Spain}

\begin{abstract}
Be/X-ray binaries are the most numerous class of X-ray binaries. They
constitute an excellent tracer of star formation and can be used to
study several aspects of astrophysics, from mass loss in massive stars
to binary evolution. This short review, intended for the
non-specialist, presents a summary of 
their basic observational properties and outlines the physical
mechanisms giving rise to 
these characteristics.
\end{abstract}
\thispagestyle{plain}

\section{What are we calling a Be/X-ray binary?}

A Be/X-ray binary can be trivially defined as a binary system
containing a Be star, which, for some reason, produces X-ray
emission. Modern reviews of the properties of Be/X-ray binaries can
be found in  \citet[concentrating on X-ray properties]{nag01} and
\citet[mainly optical observations]{coe00}. A Be star is trivially
defined as ``a non-supergiant B-type 
star whose spectrum has, or had at some time, one or more Balmer
lines in emission'' \citep{col87}. However, such trivial definitions
are necessarily too broad. If we want to define a class of
objects with common physical characteristics, these definitions need
some qualification. 

For a start, we concentrate on {\it ``classical Be stars''}, 
early-type (mostly B-type, but also late O-type)
stars which show emission lines because they are surrounded by a disk
of material lost from their equator (see \citealt{pr03} for a recent
review; see also \citealt{bal00}; \citealt{sle88}). The mass loss in a
classical Be star is due to causes 
intrinsic to the star itself (though binary companions, when present,
may have some 
triggering effect; cf.~\citealt{mir03}). In a Be/X-ray binary,
emission lines should be associated with a classical Be star and come
from such a {\it decretion disk} \citep{oka01}. A system like the black hole
candidate LMC X-3 \citep{cow83} is not a Be/X-ray binary, as the
emission lines most likely come from an accretion disk around the
black hole. The case of the
bright transient A\,0538$-$66 is less clear, as it looks like a Be star
during quiescence states, but has a spectrum completely different from
a Be star when in outburst \citep{cha83} and displays optical
variability unprecedented in a Be star \citep{mgc03}.

At present, we know the optical counterparts of $>20$ Be/X-ray
binaries in the Galaxy and $>10$ in the Large Magellanic Cloud
(LMC). A relatively up-to-date list of massive X-ray binaries, with
their properties, is given by \citet{liu00} and a recent list of
Be/X-ray binaries and candidate is provided by \citet{pr04}. All 
the counterparts have spectral type earlier than B2 \citep{neg98}. As
a matter of 
fact, the spectral distribution of the counterparts is very strongly
peaked around spectral types B0-B0.5 \citep{nc02}, suggesting that
they all have similar masses. 

It is important to note that
isolated Be stars do also display X-ray 
emission. Early-type ($<$B2) stars in general show X-ray
emission with $L_{\rm X}\sim 10^{-7} L_{\rm bol}$ (see
\citealt{ber97}; see also \citealt{har79}; \citealt{pal81}) and Be
stars may be marginally brighter \citep{coh00}. In order to have an
X-ray binary, the main X-ray source must not be the Be
star, but a binary companion, specifically a {\it compact
  companion}: a white dwarf, neutron star or black hole (a general
review of the properties of X-ray binaries can be found in
\citealt{wnp95}).  

The compact companion has no
immediate observational signatures apart from the X-ray emission. The
optical/infrared flux is completely dominated by the Be star
\citep{vpm95}. This 
results in a fundamental observational bias: an object is recognised
as a Be/X-ray binary because it
shows an X-ray flux higher than expected for an isolated Be star of
its spectral type. Considering the uncertainties in the X-ray flux
expected, and more importantly, in the distance and reddening derived to a
single star, it is relatively difficult to establish whether a given
star fulfils this criterion. Objects displaying an $L_{\rm X}$ much
higher than an isolated Be star are readily identified as Be/X-ray
binaries, while objects only one or two orders of magnitude brighter
than an isolated Be star fall into a ``grey zone'', where their binary
nature is difficult to ascertain. Because of this, the population of
objects well studied -- and even the population of objects known -- is
strongly biased toward high $L_{\rm X}$ sources, even if they show up
as such only sporadically.

If we can accumulate enough photons, we can always look for
signatures of the compact companion in the X-rays, such as a
characteristic X-ray spectrum, or 
X-ray pulsations. With sufficient monitoring, tell-tale variability may
be detected. A determination of $\dot{P}_{{\rm spin}}$ (which, of
course requires observations over a few years) may allow the
differentiation between a neutron star and a white dwarf. So with
unlimited observing time on very sensitive X-ray 
telescopes, the observational bias toward high $L_{\rm X}$ systems
could perhaps be removed, but, as we stand, it is very obviously
present and, for many weak X-ray sources, we simply do not know if
they are Be/X-ray binaries or not.

So far, all Be/X-ray binaries that have been observed with sufficient
 sensitivity have revealed the signatures of a neutron star. Indeed,
 X-ray pulsations have always been found, the only exception being the
 microquasar  1E\,0236.6+6100 (see below).

A final question to consider is the origin of the X-ray
emission. Traditionally, one talks of X-ray binaries when the physical
mechanism producing the X-rays is accretion on to a compact
object. There are cases, however, when other sources of energy are
available. One clear example is the
 radio pulsar PSR~B1259$-$63, which orbits the Be star LS~2883
 \citep{joh92}. The  
 neutron star is young and powered by dissipation of rotational
 energy. X-rays are believed to originate in shocks
 at the interface between the pulsar wind and the disk of the Be star
 \citep[][and references therein]{mur03}. Another system that could be
 powered by rotational energy is the 34-ms pulsar SAX~J0635+0533
 \citep{cus00}. The microquasar 1E\,0236.6+6100 could be similar to
 PSR~B1259$-$63 \citep{mt81}, though an accretion-powered source is currently
 favoured \citep{mas04}. All these objects have properties widely
 differing from 
 those of the majority of Be/X-ray binaries and will thus be excluded from
 the following, where we concentrate on systems containing {\it an
   X-ray pulsar accreting from the disk of a classical Be star.}

\begin{centering}
\begin{figure}
\plotfiddle{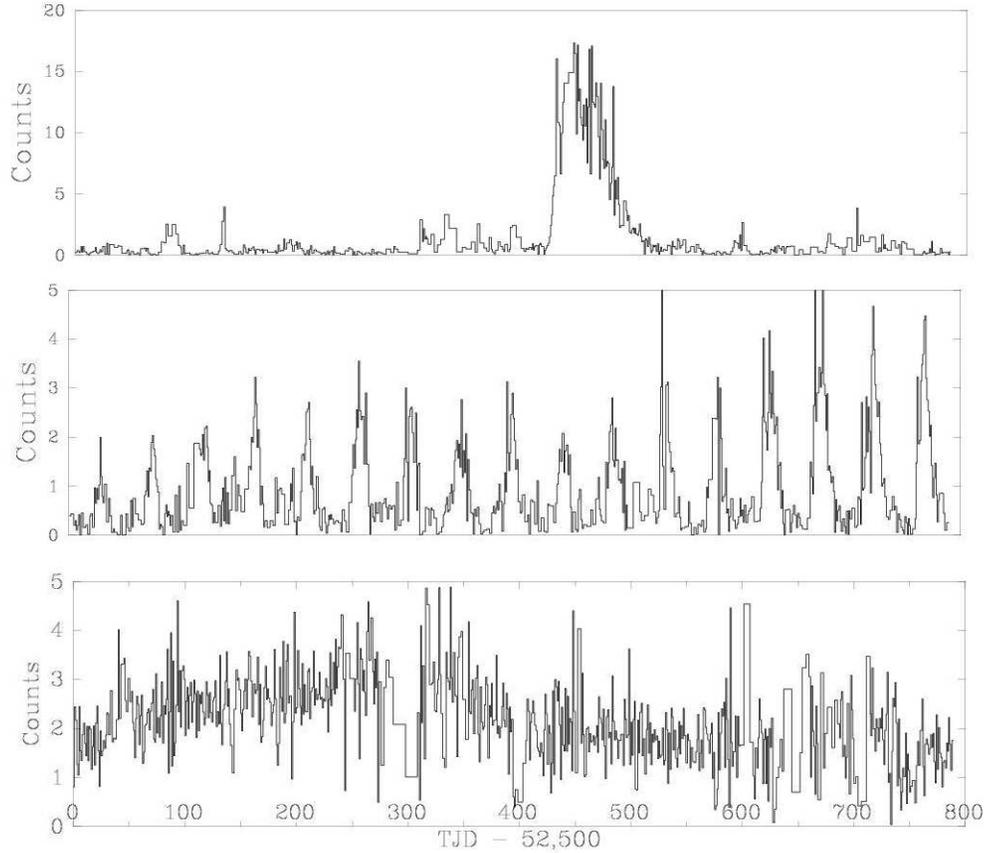}{12cm}{0}{50}{50}{-180}{0}
\caption{X-ray lightcurves of representative Be/X-ray binaries, from
  the All Sky Monitor on board {\it RossiXTE}, spanning 800 days.
{\bf  Bottom panel.} Lightcurve of 
  the prototype persistent source X Persei. The flux is always
  different from zero and varies smoothly. Sharp peaks are mostly
  associated with low signal-to-noise points or solar
  contamination. {\bf Mid panel.} Lightcurve of EXO\,2030+375, showing
  a long series of Type I outbursts, close to the time of
  periastron. {\bf Top panel.} The Be/X-ray transient X0656$-$072
  displays a single Type II outburst after $\sim 30$~years of
  inactivity. The scale of the vertical axis in the
  different panels is very different, allowing a estimation of the
  level of variability. X~Persei is the  Be/X-ray binary closest to the
  Sun known and its flux varies between $\sim   
10^{34}-10^{35}\:{\rm erg}\,{\rm s}^{-1}$. The Type~I outbursts in
  EXO\,2030+375 generally reach $\sim10^{37}\:{\rm erg}\,{\rm
  s}^{-1}$. The distance to X0656$-$072 has not been determined yet,
  but generally Type~II outbursts reach close to $\sim10^{38}\:{\rm erg}\,{\rm
  s}^{-1}$.  } 
\end{figure}
  \end{centering}

\section{X-ray properties}

As mentioned, all Be/X-ray binaries, when observed with
sufficient sensitivity, display X-ray pulsations, a signature of the
strong magnetic field ($B\sim10^{12}\:$G) of a neutron star. The presence of
X-ray pulsations allows the determination of the orbital parameters of
the system, such as the orbital period, $P_{\rm orb}$, and
eccentricity, $e$ \citep[e.g.,][]{rap78,fin99}. 
The X-ray spectra of Be/X-ray binaries are very similar to those of
other accreting X-ray pulsars, as they depend mostly on the
physical conditions close to the 
neutron star \citep[cf.][]{bil97}. They can generally be characterised by
broken power laws, with a high-energy cutoff and absorption at low
energies due to interstellar material \citep{wnp95,nag01}. In a few
systems with low interstellar absorption, there is evidence for a soft
blackbody component at low energies \citep{nag01}.

The first Be/X-ray binaries were identified as bright X-ray transient
sources \citep{mar75}, but, as new systems were discovered, very
different behaviours were observed.
Some Be/X-ray binaries are persistent X-ray sources
\citep[see][]{rr99}, displaying low luminosity ($L_{{\rm x}} \sim   
10^{34}\:{\rm erg}\,{\rm s}^{-1}$) at a relatively constant level
(varying by up to a factor of $\sim 10$). On the other hand, most
known Be/X-ray binaries (though this is likely a selection effect)
undergo outbursts in which the X-ray luminosity suddenly increases by a
factor $\ga 10$.
A given transient can show one or both of the two kinds of outbursts
\citep[cf.][]{swr86}: 

\begin{itemize}
\item X-ray outbursts of moderate intensity ($L_{{\rm x}} 
   \sim 10^{36}$ erg s$^{-1}$) occurring in series
   separated by the orbital period (Type I or normal), generally close to
   the time of periastron passage of the neutron 
   star. In most systems, the duration of normal outbursts seems
   related to the orbital period.
\item Giant (or Type II) X-ray outbursts ($L_{{\rm x}} 
   \ga 10^{37}$ erg s$^{-1}$) lasting for several weeks or even
   months. Generally, Type II outbursts start shortly after periastron
   passage, but do not show any other correlation with orbital
   parameters \citep{fp97}. 
In some systems, the duration of the Type II outbursts seems to 
be correlated with their peak intensity, but this is not always the
case \citep[cf.][]{fin96,mot91}.
\end{itemize}

During giant outbursts, and sometimes during normal outbursts, the
spin period of the neutron star is observed to increase (neutron star
{\it spin-up}), indicating that angular 
momentum is efficiently transfered 
from the material accreted to the neutron star, most likely through an
accretion disk \citep[e.g.,][]{fin99,wil03}.

The X-ray spectra of persistent sources show some differences
with respect to those of transients in outburst \citep{rr99}. Moreover, all
persistent sources have relatively long pulse periods, 
  $>200\:$s, while transients have periods ranging from less than one
second to several hundred seconds.

Be/X-ray binaries fall within a narrow area in the 
$P_{{\rm orb}}$/$P_{{\rm spin}}$ diagram \citep[see][]{cor86}. This
correlation between $P_{{\rm orb}}$ and $P_{{\rm spin}}$ is generally
interpreted as meaning that the neutron stars in Be/X-ray binaries
rotate at the equilibrium velocity between the spin-up caused by
accreted matter and the spin down caused by the centrifugal effect of
their strong magnetic fields \citep{wk89}. The correlation is loose,
and there are some clear exceptions (for instance, X0726$-$260;
\citealt{cp97}).  

\section{Optical/infrared properties}
As mentioned above, the optical/infrared properties of Be/X-ray
binaries are those of the Be star, and so very similar to those of
isolated Be stars: emission in the Balmer lines and some
singly-ionised metallic lines, infilling or emission
in the He\,{\sc i} lines and an infrared excess with respect to B-type
stars of the same spectral type, resulting in photometric
variability \citep[cf.][]{pr03,sle88}.  

In Be/X-ray binaries, the maximum strength of
H$\alpha$ ever measured correlates with the size of the orbit,
measured through $P_{\rm orb}$ \citep{rei97}. This is understood as a
consequence of the interaction between the neutron star and the disk
of the Be star. In the truncated viscous disk model \citep{on01}, the
tidal torque of the neutron star truncates the disk at the resonances
between the orbital periods of disk particles and the neutron star. As
a consequence, material accumulates in the disk, explaining why the
disks of Be/X-ray binaries appear denser than those of isolated Be
stars \citep{zam01}. This situation is necessarily unstable and will
eventually lead to major perturbations in the disk structure. Such
perturbations will result in the onset of the giant outbursts
\citep{neg01}.

In several systems, we observe relatively quick (a few years)
quasi-periodic cycles, during which the disk forms, grows, gives rise
to X-ray activity and then disappears \citep[e.g.,][]{rei01,hai04}. In
the well-studied system 4U\,0115+63, these quasi-cycles are highly
repeatable \citep{neg01}. As the mechanisms involved are rather
complex, the correlation between the optical and infrared lightcurves
and the X-ray lightcurves are rather loose
\citep[e.g.,][]{cla99}. Similar quasicycles are observed in isolated
Be stars, though they tend to last longer than in Be/X-ray binaries
\citep[e.g.,][]{cla03}. 

The size of the truncated disk depends strongly on the
orbital parameters of the system, notably the semi-major axis and
$e$. If $e$ is large,  
truncation is not very effective and the Be/X-ray binary is expected
to display Type I outbursts at every periastron passage (similarly to
EXO\,2030+375 in Fig.~1). If $e$ is low, truncation is very effective
and activity should be rare. In intermediate cases, more complex
behaviour is expected \citep{on01,oka02}. 

X~Per, the prototypical persistent source, is
known to have a wide, low-$e$ orbit \citep{del01}. Because of their
long $P_{\rm 
  spin}$ and the $P_{\rm orb}$/$P_{\rm spin}$ correlation, all
{\it persistent} Be/X-ray binaries are believed to have similar orbits. The
existence of Be/X-ray binaries  with both low and high values of $e$
has a bearing on models for their formation \citep{vhvp,pfa02} and may
even have
implications for our understanding of supernovae \citep{pod04}. 

\section{The population of Be/X-ray binaries}

Be/X-ray binaries are thought to be the product of the evolution of a
binary containing two moderately massive stars, which undergoes
mass transfer from the originally more massive star on to its companion
\citep[see][]{pol91,vvh95,vv97}. As such, they are
necessarily young and trace recent star formation.

Be/X-ray binaries are very numerous. Extrapolations from the observed
numbers suggest that there are a few thousands of them in the Galaxy
\citep{vpm95}, while
estimates based on population synthesis models predict that the number
of B-type stars with a neutron star companion is $>$10,000
\citep{mh89}. Some authors have assumed that the mass transfer phase
leading to the formation of the Be/X-ray binary necessarily forces the
B-type companion of a neutron star to be a Be star. This is at present
not an obvious conclusion (see discussion in \citealt{vv97}) and it
may well be that the majority of these systems can never be seen as
X-ray sources: if the B-type star is not in a Be phase, the neutron
star has nothing to accrete.

A major discrepancy between population synthesis models and
observations are the relative numbers 
of Be/X-ray binaries (Be + neutron star) and their lower $L_{\rm X}$
relatives, the Be + white dwarf (wd) binaries. All models predict very
large numbers of Be+wd systems, in most cases outnumbering 
Be/X-ray binaries by a factor $>10$
\citep[e.g.,][]{vv97,rag01}. Unfortunately, though there are a few
candidates to be Be+wd binaries \citep{mot97,to01}, so far no system has been
unambiguously 
confirmed to be a Be+wd binary by observations. 

There are strong selection effects against the detection of Be+wd
binaries. For a start, their expected $L_{\rm X}$ is not much higher
than that of an isolated Be star. Their relatively soft X-ray
spectra are strongly affected by interstellar absorption, meaning that
it becomes difficult to differentiate them from weak Be/X-ray binaries
with neutron stars in wide orbits (and hence slow
pulsations). In spite of this, because of their large numbers, we
should expect 
to have found some in our immediate neighbourhood, but searches for them
have so far failed \citep{meu92}, rendering population synthesis
models somewhat suspect.

\section{The SMC: laboratory for the Be/X-ray binaries} 

For the last few years, observations with a new generation of X-ray
telescopes offering good spatial resolution have revealed the presence
of a huge population of Be/X-ray binaries in the Small Magellanic
Cloud (SMC; cf.~\citealt{hs00,yok03}). To date, there are close to 40
X-ray pulsars in the SMC \citep{hp04}. All of them, except SMC X-1,
are Be/X-ray binaries. Such large  
population of objects at a given distance, with similar chemical composition,
and very little affected by 
interstellar absorption renders the SMC the perfect laboratory to
study Be/X-ray binaries.

The main limitation of the SMC is its large distance. Dedicated
pointings with X-ray telescopes are needed to detect X-ray pulsations
and existing and previous all sky monitors do not detect or resolve
its sources. For this reason, knowledge of the orbital parameters of
SMC Be/X-ray binaries is almost null. However, some promising
techniques are being developed. \citet{maj04} have shown that, because
of the $P_{\rm orb}$/$P_{\rm spin}$ correlation, there is a good
statistical correlation between the maximum observed $L_{\rm X}$ and
$P_{\rm spin}$, which can hence be used as a measurement of orbital
size. Moreover, \citet{lay04} show how $P_{\rm orb}$ can sometimes be
derived 
from observations with non-imaging instruments.

Among the first fruits of work on the large SMC
sample of Be/X-ray binaries, \citet{lay04} show an excellent correlation
between the distribution of Be/X-ray binaries and star-forming regions
in the SMC, a promising result for the study of more distant
galaxies. Meanwhile, \citet{coe04} find that the spectral distribution
of counterparts to SMC Be/X-ray binaries is very different from that
of Milky Way systems, seriously challenging many current models.

\acknowledgments{ 
The author is a researcher of the
programme {\em Ram\'on y Cajal}, funded by the Spanish Ministerio de
Ciencia y Tecnolog\'{\i}a and the University of Alicante.
This research is partially supported by the Spanish MCyT under grant
AYA2002-00814. I thank Pere Blay for his help in producing the figure,
which uses quick-look results provided by the ASM/RXTE team. I also
would like to thank Simon Clark and Christian Motch for careful reading
of the manuscript and very helpful comments.}


\begin{thebibliography}{}
\bibitem[Balona(2000)]{bal00} 
Balona, L.A. 2000,
In: Smith, M., Henrichs, H.F., \& Fabregat, J.\ (eds.)
      IAU Colloq.\ 175,
      The Be Phenomenon in Early-Type Stars.
      ASP, San Francisco, p. 1
\bibitem[Bergh\"ofer et al.(1997)]{ber97}
Bergh\"ofer, T.W., et al. 1997, A\&A 322, 167
\bibitem[Bildsten et al.(1997)]{bil97}
Bildsten, L., et al. 1997, ApJS 113, 367
\bibitem[Charles et al.(1983)]{cha83} 
Charles, P.A., et al. 1983, MNRAS 202, 657
\bibitem[Clark et al.(1999)]{cla99}
Clark, J.S., et al. 1999, MNRAS 302, 167
\bibitem[Clark et al.(2003)Clark, Tarasov \& Panko(2003)]{cla03}
Clark, J.S., Tarasov, A.E., \& Panko, E.A. 2003, A\&A 403, 239
\bibitem[Coe(2000)]{coe00} 
Coe, M.J. 2000,
In: Smith, M., Henrichs, H.F., \& Fabregat, J.\ (eds.)
      IAU Colloq.\ 175,
      The Be Phenomenon in Early-Type Stars.
      ASP, San Francisco, p. 656
\bibitem[Coe et al.(2004)]{coe04}
Coe, M.J., et al. 2004, MNRAS, in press ({\tt astro-ph/0410074})
\bibitem[Collins(1987)]{col87}
Collins, G.W.H. 1987, 
In Slettebak, A., \& Snow, T.P.\ (eds.)
      IAU Colloq.\ 92,
      Physics of Be Stars.
      CUP, Cambridge, p. 3
\bibitem[Cohen(2000)]{coh00} 
Cohen, D.H. 2000,
In: Smith, M., Henrichs, H.F., \& Fabregat, J.\ (eds.)
      IAU Colloq.\ 175,
      The Be Phenomenon in Early-Type Stars.
      ASP, San Francisco, p. 156
\bibitem[Corbet(1986)]{cor86} 
Corbet, R.H.D. 1986, MNRAS 220, 1047
\bibitem[Corbet \& Peele(1997)]{cp97} 
Corbet, R.H.D., Peele, A.G. 1997, ApJ 489, L83
\bibitem[Cowley et al.(1983)]{cow83} 
Cowley, A.P., et al. 1983, ApJ 272, 118
\bibitem[Cusumano et al.(2000)]{cus00}
Cusumano, G., et al. 2000, ApJ 528, L25
\bibitem[Delgado-Mart\'{\i} et al.(2001)]{del01} 
Delgado-Mart\'{\i}, H., et al. 2001, ApJ 546, 455
\bibitem[Finger \& Prince(1997)]{fp97} 
Finger, M. H., \& Prince, T. A. 1997, 
In Proceedings of the Fourth Compton Symposium, AIP, Woodbury, NY,
part 1, p. 57 
\bibitem[Finger et al.(1996)Finger, Wilson \& Harmon(1996)]{fin96}
  Finger, M. H., Wilson, R. B., \&  Harmon, B. A. 1996, ApJ 459, 288
\bibitem[Finger et al.(1999)]{fin99}
Finger, M.H., et al. 1999, ApJ 517, 449
\bibitem[Haberl \& Sasaki(2000)]{hs00}
Haberl, F., \& Sasaki, M. 2000, A\&A 359, 573
\bibitem[Haberl \& Pietsch(2004)]{hp04}
Haberl, F., \& Pietsch, W. 2004, A\&A 414, 667
\bibitem[Haigh et al.(2004)Haigh, Coe \& Fabregat(2004)]{hai04}
Haigh, N.J., Coe, M.J., \& Fabregat, J. 2004, MNRAS 350, 1457
\bibitem[Harnden et al.(1979)]{har79}
Harnden F.R. Jr., et al. 1979, ApJ, 234, L51
\bibitem[van den Heuvel \& van Paradijs(1997)]{vhvp} 
van den Heuvel,  E.P.J., \& van Paradijs, J. 1997, ApJ 483, 339 
\bibitem[Johnston et al.(1992)]{joh92}
Johnston, S., et al. 1992, ApJ 387, L37
\bibitem[Laycock et al.(2004)]{lay04}
Laycock, S., et al. 2004, ApJ, in press ({\tt astro-ph/0406420})
\bibitem[Liu et al.(2000)Liu, van Paradijs \& van den Heuvel(2000)]{liu00} 
Liu, Q.Z., van Paradijs, J., \& van
  den Heuvel, E.P.J. 2000, A\&AS,147, 25 
\bibitem[Majid et al.(2004)Majid, Lamb \& Macomb(2004)]{maj04}
Majid, W.A., Lamb, R.C., \& Macomb, D.J. 2004, ApJ 609, 133
\bibitem[Maraschi \& Treves(1981)]{mt81}
Maraschi, L., \& Treves, A. 1981, MNRAS 194, P1
\bibitem[Maraschi et al.(1976)Maraschi, Treves \& van den Heuvel]{mar75}
Maraschi, L., Treves, A. \& van den Heuvel, E.P.J. 1976, Natur 259, 292
\bibitem[Massi(2004)]{mas04}
Massi, M. 2004, A\&A 422, 267
\bibitem[McGowen \& Charles(2003)]{mgc03}
McGowen, K.E., \& Charles, P.A. 2003, MNRAS 339, 748
\bibitem[Meurs \& van den Heuvel(1989)]{mh89}
Meurs, E.J.A., \& van den Heuvel, E.P.J. 1989, A\&A 226, 88
\bibitem[Meurs et al.(1992)]{meu92}
Meurs, E.J.A., et al. 1992, A\&A 265, L41
\bibitem[Miroshnichenko et al.(2003)]{mir03}
Miroshnichenko, A.S., et al. 2003, A\&A 408, 305
\bibitem[Motch et al.(1991)]{mot91}
Motch, C., et al. 1991, ApJ 369, 490
\bibitem[Motch et al.(1997)]{mot97}
Motch, C., et al. 1997, A\&A 323, 853
\bibitem[Murata et al.(2003)]{mur03}
Murata, K., et al. 2003, PASJ 55, 473
\bibitem[Nagase(2001)]{nag01}
Nagase, F. 2001,
In: White, N.E., Malaguti, G., \& Palumbo, G.G.C (eds.)
X-ray Astronomy: Stellar Endpoints, AGN and the Diffuse X-ray
Background, American Institute of Physics Conference Proceedings,
vol. 599, p. 254  
\bibitem[Negueruela(1998)]{neg98} 
Negueruela, I. 1998, A\&A 338, 505
\bibitem[Negueruela \& Coe(2002)]{nc02} 
Negueruela, I., \& Coe, M.J. 2002, A\&A 385, 517
\bibitem[Negueruela et al.(2001)]{neg01}
Negueruela, I., et al. 2001, A\&A 369, 117
\bibitem[Okazaki(2001)]{oka01}
Okazaki, A.T. 2001, PASJ 53, 119
\bibitem[Okazaki \& Negueruela(2001)]{on01}
Okazaki, A.T., \& Negueruela, I. 2001, A\&A 377, 161
\bibitem[Okazaki et al.(2002)]{oka02}
Okazaki, A.T., et al. 2002, MNRAS 337, 967
\bibitem[van Paradijs \& McClintock(1995)]{vpm95}
van Paradijs, J., \& McClintock, J.E. 1995,
In: Lewin, W.H.G, van Paradijs, J., \& van den Heuvel, E.P.J.\ (eds.),
X-ray Binaries,
Cambridge University Press, p. 58
\bibitem[Pallavicini et al.(1981)]{pal81}
Pallavicini, R., et al., 1981, ApJ, 248, 279 
\bibitem[Pfahl et al.(2002)]{pfa02} 
Pfahl, E., et al. 2002, ApJ 574, 364
\bibitem[Podsiadlowski et al.(2004)]{pod04}
Podsiadlowski, Ph., et al. 2004, ApJ 612, 1044
\bibitem[Pols et al.(1991)]{pol91} 
Pols, O.R., et al. 1991, A\&A 241, 419
\bibitem[Popov \& Raguzova(2004)]{pr04}
Popov, S.B., \& Raguzova, N.V. 2004, {\tt astro-ph/0405633}
\bibitem[Porter \& Rivinius(2003)]{pr03}
Porter, J.M., \& Rivinius, T. 2003, PASP 115, 1153
\bibitem[Raguzova(2001)]{rag01}
Raguzova, N.V. 2001, A\&A 367, 848
\bibitem[Rappaport et al.(1978)]{rap78}
Rappaport, S., et al. 1978, ApJ 224, L1
\bibitem[Reig \& Roche(1999)]{rr99} 
Reig, P., \& Roche, P. 1999,  MNRAS 306, 100 
\bibitem[Reig et al.(1997)Reig, Fabregat \& Coe(1997)]{rei97}
Reig, P., Fabregat, J., \& Coe, M.J. 1997, A\&A 322, 183
\bibitem[Reig et al.(2001)]{rei01}
Reig, P., et al. 2001, A\&A 367, 266
\bibitem[Slettebak(1988)]{sle88}
Slettebak, A. 1988, PASP 100, 770
\bibitem[Stella et al.(1986)Stella, White \& Rosner(1986)]{swr86} 
Stella, L., White, N.E., \& Rosner, R. 1986, ApJ 308, 669 
\bibitem[Tavani et al.(1994)Tavani, Arons \& Kaspi(1994)]{tav94}
Tavani, M., Arons, J., \& Kaspi, V.M. 1994, ApJ 433, L37
\bibitem[Torrej\'on \& Orr(2001)]{to01}
Torrej\'on, J.M., \& Orr, A. 2001, A\&A 377, 148
\bibitem[van Bever \& Vanbeveren(1997)]{vv97} 
van Bever, J., \&  Vanbeveren, D. 1997, A\&A 322, 116 
\bibitem[Verbundt \& van den Heuvel(1995)]{vvh95}
Verbundt, F., \& van den Heuvel, E.P.J. 1995,
In: Lewin, W.H.G, van Paradijs, J., \& van den Heuvel, E.P.J.\ (eds.),
X-ray Binaries,
Cambridge University Press, p. 457
\bibitem[Waters \& van Kerkwijk(1989)]{wk89}
Waters, L.B.F.M., \& van  Kerkwijk, M.H. 1989, A\&A, 223, 196
\bibitem[White et al.(1995)White, Nagase \& Parmar(1995)]{wnp95}
White, N.E., Nagase, F. \& Parmar, A.N. 1995,
In: Lewin, W.H.G, van Paradijs, J., \& van den Heuvel, E.P.J.\ (eds.),
X-ray Binaries,
Cambridge University Press, p. 1
\bibitem[Wilson et al.(2003)]{wil03}
Wilson, C.A., et al. 2003, ApJ 584, 996
\bibitem[Yokogawa et al.(2003)]{yok03}
Yokogawa, J., et al. 2003, PASJ 55, 161
\bibitem[Zamanov et al.(2001)]{zam01}
Zamanov, R.K., et al. 2001, A\&A 367, 884

\end{thebibliography}
\end{document}